\documentstyle[aps,amsfonts]{revtex}
\def\narrowtext{}

\def\Bold#1{{\bbox #1}}
\def\sgn{{\rm sgn}}
\def\AtSigma{|_{\scriptscriptstyle\Sigma}}
\def\OnSigma{_{\scriptscriptstyle\Sigma}}

\def\bfy{\Bold{\Theta}}
\def\Yzero{\bfy_0}
\def\ypm{\bfy^\pm}
\def\YPMzero{\ypm_0}
\def\mpm{{M^\pm}}
\def\d{\Bold{\delta}}
\def\Dzero{\d_0}
\def\Done{\d_1}

\def\Grad#1{\nabla\!_{\scriptscriptstyle #1}}
\def\GRAD#1#2{\nabla^#1_{\!\scriptscriptstyle #2}}
\def\Lie#1{\hbox{\pounds}\!_{\scriptscriptstyle #1}}
\def\ip#1{i_{\scriptscriptstyle #1}}
\def\div{{\rm div}}
\def\Disc#1{[\![#1]\!]}
\def\hat#1{\widehat{#1}}
\def\bar{\overline}

\def\PRD{Phys.\ Rev.\ {\bf D}}

\begin{document}


\title{
{\normalsize gr-qc/9701047}
\hfill{\normalsize 6 December 1996}
\\[3mm]
  {\bf TENSOR DISTRIBUTIONS IN THE PRESENCE OF DEGENERATE METRICS}
  }

\author{Tevian Dray}
\address{
Dept.\ of Physics and Mathematical Physics,
		University of Adelaide, Adelaide, SA 5005, AUSTRALIA \\
School of Physics and Chemistry, Lancaster University
		Lancaster LA1 4YB, UK \\
Department of Mathematics, Oregon State University,
		Corvallis, OR  97331, USA
\footnote{Permanent address} \\
{\tt tevian{\rm @}math.orst.edu} \\[2mm]
}

\maketitle

\widetext


\begin{abstract}
Tensor distributions and their derivatives are described without assuming the
presence of a metric.  This provides a natural framework for discussing tensor
distributions on manifolds with degenerate metrics, including in particular
metrics which change signature.
\end{abstract}

\narrowtext

\section{INTRODUCTION}

The ordinary (scalar) distributions on a manifold are the continuous linear
functionals on test functions on the manifold, and so require for their
definition only the (topological) space of test functions.  Similarly, tensor
distributions are continuous linear functionals on test tensors.  The
additional structure of a volume element allows the association of a tensor
distributions to each locally integrable tensor field.  On a manifold with
(nondegenerate) metric, it is natural to use the metric volume element for
this purpose.  Furthermore, the Levi-Civita connection can be extended to act
on tensor distributions.

In the absence of a metric, one can still associate tensor distributions to
tensor fields by introducing a (nondegenerate) volume element.  Of particular
interest is the case of a metric which is degenerate at a hypersurface.  While
much of the standard theory of tensor distributions carries over directly
using this volume element, the lack of a (global) metric-compatible connection
appears to preclude the existence of a preferred notion of covariant
differentiation of tensor distributions.  We investigate here the extent to
which the covariant derivative corresponding to the (local) metric-compatible
connection can be extended to tensor distributions.  Our main result is that
this can indeed be done for a large class of degenerate metrics.

We consider here two types of degenerate metrics, those which are {\it
discontinuous} at a hypersurface, and those which are {\it continuous} but
have vanishing determinant at a hypersurface.  In both cases, we assume that
the metric is piecewise smooth, and that the pullback of the metric to the
hypersurface is the same from both sides and nondegenerate.  These two cases
are typical of {\it signature-changing metrics}, where the orthogonal
direction to the hypersurface changes character from timelike to spacelike as
one crosses the hypersurface.  However, the formalism we construct does not
involve signature change explicitly, and is thus also applicable to degenerate
metrics which do not change signature provided they satisfy the above
conditions.

The bulk of the paper is devoted to deriving the properties of differential
operations on tensor distributions in the case where the association of tensor
distributions to tensor fields is made via an arbitrary volume element.  Our
presentation parallels that of \cite{Hartley}, which presents similar results
using elegant mathematics.  We opt here instead for a more pedagogical,
self-contained approach.  Thus, in Section~\ref{Distributions} we first review
the definition of tensor distributions and the role of the volume element,
then introduce the Heaviside and Dirac distributions and show that they have
the expected properties.  In Section~\ref{Differentiation} we define the
partial, exterior and covariant derivatives of tensor distributions.  In
Section~\ref{Structure} we briefly review currents, and discuss their relation
to the usual formulation \cite{DeWitt,Lichnerowicz,Friedlander} of tensor
distributions in the presence of a metric.

Having established our formalism, in Section~\ref{Regular} we apply it to
tensor distributions corresponding to regularly discontinuous tensors, and,
more importantly, we give a possible generalization to the case of a regularly
discontinuous connection.  In Section~\ref{WithoutMetric} we consider the
Levi-Civita connection and possible choices of a volume element in spacetimes
having a degenerate metric.  Finally, a detailed example involving a
signature-changing metric is given in Section~\ref{Example}, and we discuss
our results in Section~\ref{Discussion}.

\newpage
\section{TENSOR DISTRIBUTIONS}
\label{Distributions}

Let $M$ be a smooth ($C^\infty$) $n$-dimensional manifold without boundary.
Defining the {\it test functions} on $M$ to be the space ${\cal F}$ of smooth
(real or complex) functions with compact support on $M$,
\footnote{See e.g.\ \cite{DeWitt} or \cite{Choquet} for definitions of
(equivalent) appropriate topologies on the space of test functions.}
then the (ordinary) {\it distributions} on $M$ are the continuous linear maps
 from ${\cal F}$ to ${\Bbb R}$.  This naturally leads to the definition of {\it
test tensor fields} as being the smooth tensor fields with compact support on
$M$.
\footnote{The topology on the space of test tensors can be invariantly defined
using an auxiliary (Riemannian) metric \cite{DeWitt} or via the topology on
test functions \cite{Choquet}.}  Equivalently, the components of test tensors
in any admissible coordinate chart are test functions.  {\it Tensor
distributions} $\Bold{T}$ are then the continuous linear maps from test
tensors $U$ to ${\Bbb R}$, with the result being denoted $\Bold{T}[U]$.

The product of an ordinary distribution $\Bold{D}$ and a (smooth) tensor field
$S$ may be defined by
\begin{equation}
(S\Bold{D}) [U] = \Bold{D} [\langle S,U \rangle]
\label{ProductI}
\end{equation}
where $\langle S,U \rangle$ denotes the total contraction of $S$ with the test
tensor $U$.  Furthermore, tensor distributions can be multiplied by (smooth)
functions $f$ via
\begin{equation}
(f\Bold{T}) [U] = \Bold{T} [fU]
\end{equation}
to produce a tensor distribution of the same type as the original.  These
products extend naturally to the tensor product of tensor distributions with
(smooth) tensors via
\footnote{One can also define the product of two tensor distributions as
\begin{eqnarray*}
(\Bold{S} \times \Bold{T}) [U,W] = \Bold{S}[U] \, \Bold{T}[W]
\end{eqnarray*}
but this does {\bf not} define a tensor distribution since it acts on pairs of
tensors rather than (the tensor product of) tensors.  For example,
\begin{eqnarray*}
\Bold{S}[fU] \, \Bold{T}[W] \ne \Bold{S}[U] \, \Bold{T}[fW]
\end{eqnarray*}
for arbitrary functions $f$.}
\begin{equation}
(S \otimes \Bold{T}) [U \otimes W] = \Bold{T} [\langle S,U \rangle W]
                                   = (\Bold{T} \otimes S) [W \otimes U]
\end{equation}
This tensor product can be further generalized to the wedge
product of antisymmetric tensor distributions with antisymmetric tensors.

For example, given a local frame $\{e_a\}$ and its dual coframe
$\{\omega^a\}$, consider a distribution $\Bold{\alpha}$ which maps test vector
fields $V$ to real numbers $\Bold{\alpha}[V]$.  The {\it components}
$\Bold{\alpha}_a$ of $\Bold{\alpha}$ in this frame are ordinary distributions
defined by
\begin{equation}
\Bold{\alpha}_a [f] = \Bold{\alpha} [f e_a]
\end{equation}
for any test function $f$.  Using linearity, this yields
\begin{equation}
\Bold{\alpha}[V] = \Bold{\alpha}[V^a e_a]
                 = \Bold{\alpha}_a[V^a] = \Bold{\alpha}_a[\omega^a(V)]
\end{equation}
where $\omega^a(V)=\langle \omega^a,V \rangle$ is the contraction of
$\omega^a$ with $V$, so that
\begin{equation}
\Bold{\alpha} = \omega^a \, \Bold{\alpha}_a = \Bold{\alpha}_a \, \omega^a
\end{equation}
Thus $\Bold{\alpha}$ is a covariant tensor of degree 1, a 1-form (with
distributional components).  The contraction of a (suitably smooth) vector
field $W$ with a 1-form distribution is analogous to its contraction with a
1-form field,
\begin{equation}
\Bold{\alpha}(W) = \omega^a(W) \, \Bold{\alpha}_a
                 = W^a \Bold{\alpha}_a
\end{equation}
or equivalently
\begin{equation}
\Bold{\alpha}(W)[f] = \Bold{\alpha}[fW]
\end{equation}
The extension to general tensor distributions is straightforward.  In this
manner, any tensor distribution can be written as a sum of terms, each of
which is the product of an ordinary distribution and a tensor, and is thus a
``distribution-valued tensor''.  The tensorial type of a tensor distribution
is ``dual'' to that of the test tensors it acts on, i.e.\ contravariant
indices of the test tensors correspond to covariant indices of the tensor
distribution and vice versa.  We could in fact have used this property to
define tensor distributions in terms of their components in a given basis.  In
terms of components, the operations defined in the previous paragraph are
identical to the usual formulae for tensors.  One also sees that the tensor
product of two tensor distributions cannot be defined, because it would
involve products of the component distributions.

A {\it volume element} on $M$ is a nowhere vanishing $n$-form $\omega$ on $M$.
Given a volume element $\omega$, we can associate a tensor distribution $\hat
T$ to each locally integrable tensor field $T$ via
\begin{equation}
\hat{T}[U] = \int_M \langle T,U \rangle \, \omega
\label{Action}
\end{equation}
The tensorial type of $\hat{T}$ is the same as that of $T$.

\subsection{Heaviside and Dirac Distributions}

We now define the Heaviside and Dirac distributions.  We will assume that a
volume element $\omega$ is given on $M$, and that $N$ is a given
($n$-dimensional) submanifold of $M$.  The {\it Heaviside 0-form distribution}
$\Yzero^N$ with support on $N$ is defined by
\begin{equation}
\Yzero^N[f] = \int_{N} f \omega
\end{equation}
for test functions $f$, and the {\it Dirac 1-form distribution} $\Done^N$ with
support on $\partial N$ is defined by
\begin{equation}
\Done^N[V] = \int_{\partial N} \ip{V} \omega
\end{equation}
where $V$ is a test vector field and $\ip{V}$ denotes the interior product
(e.g.\ $\ip{V}(df)=df(V)=V(f)$ for any function $f$).

These special distributions can be multiplied not just by smooth tensors, but
by suitable locally integrable tensors.  Specifically, for any locally
integrable tensor $S$ on $N$, we can define
\begin{equation}
S \, \Yzero^N [U] = \int_{N} \langle S,U \rangle \, \omega
\end{equation}
which is essentially (\ref{ProductI}), while for any locally integrable tensor
$\bar{S}$ on $\partial N$ we can define
\begin{equation}
\bar{S} \otimes \Done^N [W \otimes V] =
  \int_{\partial N} \langle \bar{S},W \rangle \, \ip{V} \omega
\end{equation}

Suppose further that a hypersurface $\Sigma$ is given by $\{\lambda=0\}$ with
$d\lambda\ne0$.  This determines a volume element $\bar\sigma$ on $\Sigma$ via
the pullback of the {\it Leray form} $\sigma$ satisfying
\begin{equation}
\omega = d\lambda \wedge \sigma
\end{equation}
so that $\bar\sigma$ depends on the choice of $\lambda$.  We can replace
$\lambda$ by $-\lambda$ if necessary to ensure that $d\lambda(Y)>0$ for all
vector fields $Y$ which point away from $N$; this fixes the orientation of
$\bar\sigma$.
\footnote{More generally, let $\mu\ne0$ be a 1-form which is {\it normal} to
$\Sigma$ in the sense that $\mu(X)=0$ for all $X$ tangent to $\Sigma$.  A
simple and common example is $\mu=d\lambda$.  This determines a volume element
$\sigma$ on $\Sigma$ via $\omega = \mu\wedge\sigma$, which of course depends
on the choice of $\mu$.}
The {\it Dirac 0-form distribution} $\Dzero^\Sigma$ with support on $\Sigma$
is defined via
\begin{equation}
\Dzero^\Sigma[f] = \int_\Sigma f \sigma
\end{equation}
But we now have
\begin{equation}
\ip{V}\omega = d\lambda(V)\,\sigma - d\lambda\wedge\ip{V}\sigma
\end{equation}
and the last term pulls back to zero on $\Sigma$, so that if
$\Sigma=\partial N$ we have
\begin{equation}
\Done^N[V] \equiv \Dzero^{\partial N}[d\lambda(V)]
\end{equation}
or equivalently
\begin{equation}
\Done = \Dzero \, d\lambda
\end{equation}
where we have dropped the indices $N$ and $\partial N$.

\section{DIFFERENTIATION OF DISTRIBUTIONS}
\label{Differentiation}

Before discussing in turn partial, exterior, and covariant differentiation, we
first recall the definition of the {\it divergence} of a (smooth) vector field
$X$ with respect to the (smooth) volume element $\omega$, namely
\begin{equation}
\div(X) \, \omega = \Lie{X} \, \omega
\label{DivDef}
\end{equation}
where
\begin{equation}
\Lie{X} = \ip{X}d + d\ip{X}
\end{equation}
denotes Lie differentiation.  For example, in a coordinate basis with
\begin{equation}
\omega = k \, dx^1 \wedge ... \wedge dx^n
\end{equation}
we have
\begin{equation}
\div\left(X^a {\partial\over\partial x^a}\right)
  = {1\over k} \, {\partial\over\partial x^a} \big( kX^a \big)
\label{MetricDiv}
\end{equation}
In Cartesian coordinates on ${\Bbb R}^n$, for which $k=1$, we recover the
usual Euclidean expression for the divergence of a vector.  More generally,
the covariant divergence $\nabla\cdot X$ of $X$ with respect to any
torsion-free connection leaving $\omega$ invariant coincides with $\div(X)$,
$\nabla\cdot X=\div(X)$; an example is the Levi-Civita connection determined
by a metric $g$, for which $k=\sqrt{|\det{g}|}$.

We now derive some useful properties of the divergence which we will need
later.  From the identity
\begin{eqnarray}
\Lie{fX}\omega
   =  d\ip{fX}\omega
   =  d\ip{X}(f\omega)
  &=& \Lie{X}(f\omega) \nonumber\\
  &=& X(f) \, \omega + f \div(X) \, \omega
\end{eqnarray}
we recover the usual formula
\begin{equation}
\div(fX) = f\div(X) + df(X)
\label{DivID}
\end{equation}
Similarly,
\begin{eqnarray}
\Lie{X}\Lie{Y}\omega
   =  \Lie{X} \big( \div(Y)\,\omega \big)
   =  \div\big(\div(Y)X\big) \, \omega
\end{eqnarray}
and
\begin{equation}
\Lie{[X,Y]} \, \omega
  = [\Lie{X},\Lie{Y}] \, \omega
  = \Lie{X}\Lie{Y}\omega - \Lie{Y}\Lie{X}\omega
\end{equation}
so that
\begin{equation}
\div\big([X,Y]\big) = \div\big(\div(Y)X\big) - \div\big(\div(X)Y\big)
\label{DivIDTwo}
\end{equation}

\subsection{Partial Differentiation}

The distribution corresponding to $X(k)$, where $X$ is a vector field and $k$
is a (locally integrable and differentiable) function on $M$ is
\begin{eqnarray}
\widehat{X(k)}[f] = \int_M X(k) f \, \omega
  &=& \int_M \Lie{fX}(k) \, \omega \nonumber\\
  &=& \int_M \Big( \Lie{fX}(k\omega) - k\Lie{fX}\omega \Big) \nonumber\\
  &=& \int_M \Big( d \ip{fX} (k\omega) - k \, \div(fX) \, \omega \Big)
   =  \int_{\partial M} k \, \ip{fX} \omega - \hat{k}[\div(fX)]
\label{PartialOne}
\end{eqnarray}
where $d\omega=0$ was used and the surface term vanishes because $f$ has
compact support.  This motivates the definition~%
\footnote{Both $X$ and $\omega$ must be at least $C^{k+1}$ if $D$ is a
distribution of order $k$.  Unless explicitly stated, all tensor fields,
including $\omega$, are assumed to be smooth ($C^\infty$).}
(which is equivalent to that of \cite{Choquet})
\begin{equation}
X(\Bold{D})[f] = -\Bold{D}[\div{(fX)}]
\label{Partial}
\end{equation}
for the action of $X$ on an ordinary distribution $\Bold{D}$.  Using the
identities (\ref{DivID}) and (\ref{DivIDTwo}) it is easy to check that the
definition (\ref{Partial}) is linear in $X$ and $\Bold{D}$, respects the
action of vector fields on functions
\begin{equation}
X(\hat{k}) = \widehat{X(k)}
\end{equation}
satisfies the product rule
\begin{equation}
X(k\Bold{D}) = X(k)\Bold{D} + kX(\Bold{D})
\end{equation}
and preserves the commutator algebra of vector fields
\begin{equation}
X \big( Y(\Bold{D}) \big) - Y \big( X(\Bold{D}) \big) = [X,Y](\Bold{D})
\end{equation}
In Cartesian coordinates on ${\Bbb R}^n$ we recover the usual expression
\begin{equation}
\left({\partial\over\partial x^a}\Bold{D}\right)[f]
  = - \Bold{D}\left[{\partial\over\partial x^a} f\right]
\end{equation}
or equivalently
\begin{equation}
X(\Bold{D})[f]
  = - \Bold{D}\left[{\partial\over\partial x^a}(fX^a)\right]
\end{equation}

For example, with $\Sigma=\partial N$ as above and $\partial_\lambda$ defined
by $\partial_\lambda(\lambda)=1$ and e.g.\ $i_{\partial_\lambda}\sigma=0$, we
have
\begin{equation}
\div(f\partial_\lambda)\,\omega = d i_{f\partial_\lambda} \omega
                                = d (f \sigma)
\end{equation}
so that
\begin{eqnarray}
\partial_\lambda(\Yzero^N) [f]
  &=& -\Yzero^N [\div(f\partial_\lambda)] \nonumber\\
  &=& -\int_N d(f\sigma) \nonumber\\
  &=& -\int_{\partial N} f\sigma
  = -\Dzero^{\partial N} [f]
\end{eqnarray}
or in other words
\begin{equation}
\partial_\lambda(\Yzero^N) = -\Dzero^{\partial N}
\end{equation}
More generally,
\begin{equation}
X(\Yzero^N) = -\Dzero^{\partial N} \, d\lambda(X) = -\ip{X}\Done^{\partial N}
\end{equation}
where
\begin{equation}
\ip{X}\Bold{\alpha} [U] = p \, \Bold{\alpha}[X \otimes U]
\end{equation}
defines the interior product acting on $p$-form distributions.  We can
continue in this manner to define further partial derivatives of these
distributions.  For instance, defining
\begin{equation}
\Dzero' = \partial_\lambda(\Dzero)
\end{equation}
leads to
\begin{equation}
\Dzero'[f] = -\Dzero[\div(f\partial_\lambda)]
\end{equation}
where we have again dropped the labels $N$ and $\partial N$.  In the special
case where $d\sigma=0$, i.e.\ $\sigma$ is independent of $\lambda$, then
$\div(f\partial_\lambda)=\partial_\lambda(f)$ and we recover the familiar
result
\begin{equation}
\Dzero'[f] = -\int_\Sigma \partial_\lambda(f) \, \sigma
           = -\Dzero[f']
\end{equation}
where $f'=\partial_\lambda(f)$.  Note that distributions are not necessarily
infinitely differentiable; their differentiability is limited by that of
$\omega$.

\subsection{Exterior Differentiation}

Antisymmetric tensor distributions of degree $(0,p)$ will be called {\it
$p$-form distributions}.  It is customary to write the action of $p$-forms
$\alpha$ on $p$ vector fields $\{X^i,i=1...p\}$ as $\alpha(X^1,...,X^p)$,
where of course
\begin{equation}
\alpha(X^1,...,X^p) = \langle \alpha , X^1 \otimes ... \otimes X^p \rangle
\end{equation}
The volume element $\omega$ provides us through (\ref{Action}) with an action
of $p$-forms $\alpha$ on $p$ vector fields $\{X^i,i=1...p\}$ as
\begin{equation}
\hat\alpha [X^1 \otimes ... \otimes X^p] = \int_M \alpha(X^1,...,X^p) \, \omega
\end{equation}
which makes $\hat\alpha$ a $p$-form distribution.

For the special case where $\alpha$ is a 1-form $df$, we note
the identity
\begin{eqnarray}
\widehat{\,df\,}\![X] = \int_M df(X) \, \omega
  &=& \int_M \ip{X}(df) \, \omega \nonumber\\
  &=& \int_M \Lie{X}(f) \, \omega \nonumber\\
  &=& \int_M \Big( \Lie{X}(f\omega) - f\Lie{X}\omega \Big) \nonumber\\
  &=& \int_M \Big( d \ip{X} (f\omega) - f \div(X) \, \omega \Big)
   =  \int_{\partial M} f \, \ip{X} \omega   - \hat{f}[\div(X)]
\label{Div}
\end{eqnarray}
and the surface term vanishes since the test vector field $X$ is supported
away from $\partial M$.  This motivates the definition
\begin{equation}
d\Bold{F}[V] = - \Bold{F} [\div(V)]
\label{Exterior}
\end{equation}
for (ordinary) distributions $\Bold{F}$, which satisfies
\begin{equation}
d\hat{f} = \widehat{\,df\,}
\end{equation}
Equivalently, we can define exterior differentiation of (ordinary)
distributions $\Bold{F}$ using (\ref{Partial}) via
\begin{equation}
d\Bold{F}(X) = X(\Bold{F})
\label{Elegant}
\end{equation}
Thus, introducing a local frame $\{e_a\}$ and its dual coframe $\{\omega^a\}$
as before
\begin{equation}
d\Bold{F}[V] = d\Bold{F}[V^a e_a]
             = d\Bold{F}(e_a)[V^a] = e_a(d\Bold{F})[V^a]
             = -\Bold{F}[\div(V^a e_a)] = -\Bold{F}[\div(V)]
\end{equation}
which agrees with (\ref{Exterior}).  The exterior calculus of $p$-form
distributions can therefore be built up in complete analogy with the exterior
calculus of $p$-forms by regarding tensor distributions as distribution-valued
tensors.  For instance, in a coordinate basis, any $p$-form distribution can
be written as a sum of terms of the form $\Bold{F} \, dx^1 \wedge ... \wedge
dx^p$, whose exterior derivative can then be defined to be
\begin{equation}
d(\Bold{F} \, dx^1 \wedge ... \wedge dx^p)
  = d\Bold{F} \wedge dx^1 \wedge ... \wedge dx^p
\label{ElegantII}
\end{equation}

In any case, applying (\ref{Exterior}) to the Heaviside distribution, we have
\begin{eqnarray}
d\Yzero^N[V] = -\Yzero^N[\div(V)]
  &=& - \int_N \div(V) \, \omega \nonumber\\
  &=& - \int_N d \ip{V} \omega \nonumber\\
  &=& - \int_{\partial N} \ip{V} \omega
   =  -\Done^N[V]
\end{eqnarray}
or in other words
\begin{equation}
d\Yzero^N = -\Done^N
\end{equation}

The formula (\ref{Exterior}) generalizes recursively to higher degree forms as
follows.  If $\alpha$ is a $p$-form, the analog of (\ref{Div}) is
\begin{eqnarray}
(p+1) \, \widehat{d\alpha}[X \otimes U]
   =  (p+1) \, \int_M \langle d\alpha, X \otimes U \rangle \, \omega
  &=& \int_M \langle \ip{X}d\alpha , U \rangle \, \omega \nonumber\\
  &=& \int_M \Big( \langle \Lie{X}\alpha ,U \rangle
                 - \langle d\ip{X}\alpha , U \rangle \Big) \, \omega
      \nonumber\\
  &=& \int_M \Big( \Lie{X} \big( \langle \alpha,U \rangle \, \omega \big)
                   - \langle \alpha , \Lie{X}U \rangle \, \omega
                   - \langle \alpha,U \rangle\Lie{X}\omega
                   - \langle d\ip{X}\alpha, U \rangle \, \omega \Big)
      \nonumber\\
  &=& \int_M d \big( \langle \alpha,U \rangle \, \ip{X} \omega \big)
       - \hat\alpha \Big[ \div(X)U + \Lie{X}U \Big] - \widehat{d\ip{X}\alpha}[U]
\label{DivTwo}
\end{eqnarray}
where $X$ is a test vector field and $U$ is a test tensor field of type
$(p,0)$, and where the surface term vanishes since $U$ is a test tensor.  This
motivates the following recursive definition
\begin{equation}
(p+1) \, d\Bold{\alpha}[X \otimes U]
  = - \Bold{\alpha} \Big[ \div(X)U + \Lie{X}U \Big]
    - d\ip{X}\Bold{\alpha}[U]
\label{ExteriorTwo}
\end{equation}
for the $p$-form distribution $\Bold{\alpha}$.  This equation can
alternatively be derived from the more elegant approach starting from
(\ref{Elegant}) and (\ref{ElegantII}).  In any case, (\ref{ExteriorTwo}) again
satisfies the compatibility condition
\begin{equation}
d\hat\alpha = \widehat{d\alpha}
\end{equation}
In the special case when $\Bold{\alpha}$ is a 1-form distribution, we have
\begin{equation}
2 \, d\Bold{\alpha}[X \otimes Y]
  = - \Bold{\alpha} \Big[ \div(X) Y - \div(Y) X + [X,Y] \Big]
\end{equation}

The usual properties of exterior differentiation, such as $d^2=0$ and the
product rule, follow directly from (\ref{Elegant}) and~(\ref{ElegantII}).
They can also be verified directly using the identities (\ref{DivID}) and
(\ref{DivIDTwo}).  For instance, for any (ordinary) distribution $\Bold{D}$ we
have
\begin{eqnarray}
2 \, d^2\Bold{D} [X \otimes Y]
  &=& -d\Bold{D} \Big[ \div(X)Y-\div(Y)X+[X,Y] \Big] \nonumber\\
  &=& \Bold{D} \Big[ \div\big(\div(X)Y\big) - \div\big(\div(Y)X\big)
                     + \div\big([X,Y]\big) \big]
   =  0
\end{eqnarray}
and
\begin{eqnarray}
d(f\Bold{D}) [X]
   =  - f\Bold{D} [\div(X)]
  &=& - \Bold{D} \big[f\div(X)\big] \nonumber\\
  &=& - \Bold{D} \big[\div(fX) - df(X)\big] \nonumber\\
  &=& d\Bold{D}[fX] + \Bold{D}[df(X)]
\end{eqnarray}
so that
\begin{equation}
d(f\Bold{D}) = f d\Bold{D} + df\Bold{D}
\end{equation}
as expected.  The product rule
\begin{equation}
d(\Bold{D}\alpha) = d\Bold{D} \wedge \alpha + \Bold{D}\,d\alpha
\label{dID}
\end{equation}
can then be established by induction.  Since as previously noted any tensor
distribution can be written as a sum of tensor products of tensors with
ordinary distributions, we have in fact established the full product rule,
namely
\begin{equation}
d\big(\alpha\wedge\Bold{\Gamma}\big) = d\alpha\wedge\Bold{\Gamma}
                                      + (-1)^p \alpha\wedge d\Bold{\Gamma}
\end{equation}
for smooth $p$-forms $\alpha$ and $q$-form distributions $\Bold{\Gamma}$.

\subsection{Covariant Differentiation}

Now suppose that not only a volume element $\omega$ is given on $M$, but also
a connection $\nabla$ (i.e.\ a connection whose covariant derivative operator
is $\nabla$).  The standard way to extend $\nabla$ to act on tensor
distributions is simply to work with tensor components, Christoffel symbols,
and partial differentiation \cite{Choquet}.  Alternatively, one can construct
the ``adjoint'' of $\nabla$ with respect to contraction, as follows.

Let $T$, $U$ be tensors of ``dual'' degree, so that the contraction
$T(U)=\langle T,U \rangle$ is defined.  We have
\begin{equation}
\Grad{X} T (U) = X \big( \langle T,U \rangle \big) - T(\Grad{X} U)
               = \Lie{X} \big( \langle T,U \rangle \big) - T(\Grad{X} U)
\end{equation}
Thus, the tensor $\Grad{X} T$, reinterpreted as a tensor distribution, acts on
the test tensor $U$ as
\begin{eqnarray}
\widehat{\Grad{X}T} [U]
  &=& \int_M \Grad{X} T (U) \, \omega \nonumber\\
  &=& \int_M \Lie{X} \big( \langle T,U \rangle \big) \, \omega
      - \hat{T} \big[ \Grad{X} U \big] \nonumber\\
  &=& \int_M \Lie{X} \big( \langle T,U \rangle \, \omega \big) 
      - \int_M \langle T,U \rangle \,\Lie{X} \omega  
      - \hat{T} \big[ \Grad{X} U \big] \nonumber\\
  &=& \int_M d \big( \langle T,U \rangle \,\ip{X} \omega \big)
      - \int_M \langle T,U \rangle \,\div(X)\,\omega 
      - \hat{T} \big[ \Grad{X} U \big] \nonumber\\
  &=& \int_{\partial M} \langle T,U \rangle \,\ip{X} \omega
      -\hat{T} \Big[ \div(X) \, U + \Grad{X} U \Big]
\label{Grad}
\end{eqnarray}
where the surface term vanishes since $U$ is a test tensor.  This motivates
the definition of the covariant derivative of any tensor distribution
$\Bold{T}$ as
\begin{equation}
\Grad{X} \Bold{T} [ U ] = -\Bold{T} \big[ \div(X) \, U + \Grad{X} U \big]
\label{Covariant}
\end{equation}
This expression agrees with the usual component definition
\cite{Lichnerowicz,Choquet} of $\Grad{X}\Bold{T}$ using partial
differentiation as given by (\ref{Partial}), which makes it clear that
$\Grad{X}$ is indeed a tensor derivation.

It is important to note that $\div(X)$ in (\ref{Covariant}) refers to the
divergence defined in (\ref{DivDef}), and not to the divergence $\nabla\cdot
X$ defined by the connection $\nabla$.  As already noted, if $\nabla$ leaves
$\omega$ invariant, these two notions of divergence agree; (\ref{Covariant})
generalizes the results of \cite{Lichnerowicz,Choquet} to the case where they
differ.  (This generalization is obtained in the component definition of the
covariant derivative by interpreting the ordinary derivatives in the sense of
equation (\ref{Partial}) above, so that they depend on the choice of
$\omega$.)

\section{MANIFOLDS WITH OTHER STRUCTURE}
\label{Structure}

\subsection{Manifolds without Volume Elements}
\label{WithoutOmega}

Even without a volume element, there is an action of $k$-forms on
$(n{-}k)$-forms, given by
\begin{equation}
\widetilde{\alpha}[\beta] = \int_M \alpha \wedge \beta
\end{equation}
which defines the {\it current} $\widetilde\alpha$ of degree $k$ \cite{de
Rham}.  Currents of degree $k$ can be uniquely associated with totally
antisymmetric {\it contravariant} tensor distributions of degree $n{-}k$.  We
can thus define the {\it Heaviside current} $\bfy^N$ with support on a
submanifold $N$ by
\begin{equation}
\bfy^N[\beta] = \int_N \beta
\end{equation}
for test $n$-forms $\beta$.  Similarly, we define the {\it Dirac current}
$\d^\Sigma$ with support on a hypersurface $\Sigma$ via
\begin{equation}
\d^\Sigma[\gamma] = \int_\Sigma \gamma
\end{equation}
for test $(n{-}1)$-forms $\gamma$.

Since test tensors vanish on the (topological) boundary of $M$, integration by
parts yields
\begin{eqnarray}
\nonumber
\widetilde{d\alpha}[\gamma] = \int_M d\alpha \wedge \gamma
  &=& \int_M d ( \alpha \wedge \gamma ) - (-1)^k \alpha \wedge d\gamma \\
  &=& 0 -(-1)^k \widetilde\alpha[d\gamma]
\end{eqnarray}
for $k$-forms $\alpha$ and $(n{-}k{-}1)$-forms $\gamma$.  We use this formula
to define exterior differentiation of currents.  For the Heaviside current, we
obtain
\begin{eqnarray}
d\bfy^N[\gamma] = -\bfy^N[d\gamma]
  &=& -\int_N d\gamma \nonumber\\
  &=& -\int_{\partial N} \gamma
   =  -\d^{\partial N}[\gamma]
\end{eqnarray}
and we thus again have a relation of the form
\begin{equation}
d\bfy^N = -\d^{\partial N}
\end{equation}

\subsection{Manifolds with Metrics}
\label{WithMetric}

If we now assume that a nondegenerate $C^0$ metric tensor $g$ is specified on
$M$, then there is a natural volume element $\omega={*}1$ on $M$, where $*$
denotes the Hodge dual determined by $g$.  This in turn determines the
Heaviside 0-form distribution $\Yzero$ and the Dirac 1-form distribution
$\Done$, which can now be expressed in terms of the Heaviside current
$\bfy$ and the Dirac current $\d$ as
\begin{eqnarray}
\Yzero[f] &=& \bfy[{*}f]
\label{MetTh} \\
\Done[V] &=& \d[{*}V^\flat]
\label{MetD}
\end{eqnarray}
where $V^\flat$ denotes the 1-form which is the metric dual of $V$ and we have
used the identity
\begin{equation}
{*}V^\flat = \ip{V}{*}1 = \ip{V}\omega
\end{equation}

So long as $\Sigma$ is not null, there is also a natural induced volume
element $\bar\sigma$ on $\Sigma$, obtained as above as the pullback of the
Leray form $\sigma$ satisfying $\omega=m\wedge\sigma$.  Here $m$ is the unique
1-form normal to $\Sigma$ having unit norm and satisfying $m(Y)>0$ for
appropriately oriented vector fields $Y$.  Normal coordinates can be used to
ensure that $m$ takes the form $m=d\lambda$.  Equivalently,
$\sigma={\bar{*}}1$, where $\bar{*}$ is the Hodge dual determined by the
pullback $h$ of $g$ to $\Sigma$; $h$ is nondegenerate since $\Sigma$ is not
null.  This enables us to determine the Dirac 0-form distribution $\Dzero$,
which can now be written
\begin{equation}
\Dzero[f] = \d[\bar{*}f]
\end{equation}
Thus, in this case there are preferred forms for the basic distributions,
arising from the preferred volume elements.

Even if $\Sigma$ is null, the volume element $\omega$ is nondegenerate, and we
can still define the distributions $\Dzero$, $\Dzero'$, and $\Done'$ provided
we are willing to choose a function $\lambda$ such that $\Sigma=\{\lambda=0\}$
as before, or equivalently provided we are willing to specify the ``induced''
volume element on $\Sigma$.  The scale freedom in this choice reflects the
lack of unit normal vector to a null surface, or equivalently the nonexistence
of normal coordinates.

As these formulas hint, the entire theory of $p$-form distributions can in
fact be elegantly rewritten in terms of yet another action of $p$-forms, this
time on $p$-forms, namely
\begin{equation}
\hat{\hat{\,\alpha\,}}[\beta] = \int_M \alpha \wedge {*}\beta
\label{HodgeAction}
\end{equation}
Using the identity
\begin{equation}
\ip{X}\alpha = {*} \left( {*}^{-1}\alpha \wedge X^\flat \right)
             = (-1)^{p+1} \,\, {*}^{-1} \! 
		\left( X^\flat \wedge {*}\alpha \right)
\label{IntProdID}
\end{equation}
it is straightforward to show that
\begin{equation}
(p!) \, {*}\Big( \alpha(X_1,...,X_p) \Big)
  \equiv \alpha \wedge {*} \left( X_1^\flat \wedge ... \wedge X_p^\flat \right)
\end{equation}
Taking ${*}^{-1}$ of both sides of this formula results in an expression for
contractions in terms of $*$, which is remarkable since it shows that the
resulting right-hand-side is thus independent of the choice of metric and
corresponding Hodge dual operator appearing in it.  This formula immediately
yields for any $p$-form $\alpha$
\begin{equation}
(p!) \, \hat{\alpha}[U] \equiv \hat{\hat{\,\alpha\,}}[U^\flat]
\end{equation}
where as usual
\begin{equation}
(X_1 \otimes ... \otimes X_p)^\flat = X_1^\flat \otimes ... \otimes X_p^\flat
\label{Flat}
\end{equation}
and where the action (\ref{HodgeAction}) has been extended by
antisymmetrization of the argument to an action on covariant tensors of type
$(0,p)$.  Explicitly,
\begin{equation}
\hat{\hat{\,\alpha\,}}[X_1^\flat \otimes ... \otimes X_p^\flat]
  = \hat{\hat{\,\alpha\,}}[X_1^\flat \wedge ... \wedge X_p^\flat]
\end{equation}
so that (\ref{HodgeAction}) is equivalent to the original action
(\ref{Action}) defined by contraction.  This equivalence allows us to
interpret $p$-form distributions $\Bold{\alpha}$ as $(n{-}p)$-currents, also
denoted $\Bold{\alpha}$, via
\begin{equation}
\Bold{\alpha} [\beta] := (p!) \, \Bold{\alpha} [\beta^\sharp]
\label{Sharp}
\end{equation}
where $\sharp$ is the inverse of (\ref{Flat}).

Since for $p$-forms $\alpha$ and $(p+1)$-forms $\beta$
\begin{eqnarray}
d(\alpha\wedge{*}\beta)
  &=& d\alpha \wedge {*}\beta + (-1)^{p+1} \alpha \wedge d{*}\beta \nonumber\\
  &=& d\alpha \wedge {*}\beta - (-1)^p \alpha \wedge {*}({*}^{-1}d{*}\beta)
\end{eqnarray}
we see that the adjoint of $d$ with respect to the action (\ref{HodgeAction})
is given by
\begin{equation}
\hat{\hat{\,d\alpha\,}}[\beta] = \hat{\hat{\,\alpha\,}}[\delta\beta]
\label{DoubleHat}
\end{equation}
where the operator $\delta$ (no relation to the tensor distribution
$\Bold{\delta}$) is defined on $p$-forms by
\begin{equation}
\delta= (-1)^p \, {*}^{-1}d{*}
\end{equation}

The action (\ref{DoubleHat}) leads to an alternative notion of exterior
differentiation of $p$-form distributions, namely
\begin{equation}
\bar{d}\Bold{\alpha}[\beta] = \Bold{\alpha}[\delta\beta]
\end{equation}
It now follows immediately that
\begin{equation}
\bar{d}^2\Bold{\alpha}[\beta]
  = \bar{d}\Bold{\alpha}[\delta\beta]
  = \Bold{\alpha}[\delta^2\beta]
  = 0
\label{Square}
\end{equation}
since $\delta^2=0$ is a consequence of $d^2=0$ and $**=\pm1$.  We next note
the identity
\begin{equation}
\Bold{\alpha} [\ip{X}\beta] = (X^\flat \wedge \Bold{\alpha}) [\beta]
\end{equation}
which follows from (\ref{IntProdID}).
\footnote{Repeated use of this equation can in fact be used to {\it define}
the exterior product of differential forms with $p$-forms.}
We further compute
\begin{eqnarray}
\bar{d}(f\Bold{\alpha})[\beta]
  &=& f\Bold{\alpha}[\delta\beta] \nonumber\\
  &=& \Bold{\alpha}[f\delta\beta] \nonumber\\
  &=& \Bold{\alpha} [\delta(f\beta) + i_{df^\sharp}\beta] \nonumber\\
  &=& \bar{d}\Bold{\alpha}[f\beta] + (df\wedge\Bold{\alpha})[\beta]
\end{eqnarray}
where $\sharp$ is the inverse of $\flat$ and where we have used the identity
\begin{equation}
\delta(f\beta) = f\delta\beta - i_{df^\sharp}\beta
\end{equation}
which again follows from (\ref{IntProdID}).  In other words
\begin{equation}
\bar{d}(f\Bold{\alpha}) = df \wedge \Bold{\alpha} + f \bar{d}\Bold{\alpha}
\label{Product}
\end{equation}

It only remains to relate the two notions of exterior derivative $d$ and
$\bar{d}$.  To avoid confusion, we think of $\bar{d}$ as acting on $p$-form
distributions via the equivalence (\ref{Sharp}), i.e.\ 
\begin{equation}
\bar{d}\Bold{\alpha}[U] = \Bold{\alpha}[(\delta U^\flat)^\sharp]
\end{equation}
We claim that
\begin{equation}
\bar{d} \equiv d
\end{equation}
This is obvious for ordinary distributions $\Bold{F}$, for which
\begin{equation}
\bar{d}\Bold{F} [X]
  = \Bold{F}[(\delta X^\flat)^\sharp]
  = \Bold{F}[\delta X^\flat]
  = -\Bold{F}[{*}^{-1}d{*}X^\flat]
  = -\Bold{F}[\div(X)]
  = d\Bold{F}[X]
\end{equation}
But since $d$ and $\bar{d}$ agree on ordinary distributions and since both
have the usual properties of exterior differentiation, they must be identical,
as claimed.
\footnote{This makes sense, since they agree on tensor distributions
arising from locally integrable tensors, and in particular from test tensors,
by virtue of (\ref{DoubleHat}), and since such tensor distributions are dense
in the space of all tensor distributions \cite{DeWitt}.  This argument could
perhaps be used to {\it prove} $d^2=0$ and the product rule from
(\ref{Square}) and (\ref{Product}).  If so, then even in the absence of a
metric one could introduce a local ``auxiliary metric'' whose metric volume
element agrees with the given volume element.  For instance, choose any
1-forms $\{\omega^i\}$ such that $\omega = \omega^1\wedge...\wedge\omega^n$
and define $\omega^i$ to be orthonormal, so that
\begin{eqnarray*}
g = \omega^1\otimes\omega^1 +...+ \omega^n\otimes\omega^n
\end{eqnarray*}
The above proofs would then hold locally, and would be independent of the
metric chosen.}

\section{REGULARLY DISCONTINUOUS TENSORS}
\label{Regular}

In many applications, tensors which are smooth on either side of a given
hypersurface play an important role.  This notion is made precise by the
notion of {\it regularly discontinuous} tensors.

We assume that we are given a hypersurface $\Sigma$ which partitions $M$ into
two disjoint open regions $\mpm$.  A tensor field $T$ is said to be {\it
regularly discontinuous} across $\Sigma$ if $T$ is continuous on $\mpm$ and
$T$ converges uniformly to tensors $T^\pm\OnSigma$ in the limit to $\Sigma$
 from $\mpm$ \cite{Lichnerowicz,Choquet}.  Note that $T$ need only be defined
on $\mpm$.  A tensor $T$ is thus regularly discontinuous if and only if its
components in any coordinate chart are regularly discontinuous functions.  The
{\it discontinuity} $\Disc{T}$ of a regularly discontinuous tensor $T$ is the
(ordinary) continuous tensor on $\Sigma$ given by
\begin{equation}
\Disc{T} = T^+\OnSigma - T^-\OnSigma
\end{equation}
More generally, we shall call $T$ {\it piecewise $C^k$} if the components of
$T$ in any coordinate chart as well as their first $k$ derivatives are all
regularly discontinuous functions.  This agrees with the notion of {\it
regularly $C^k$ discontinuous tensors} in \cite{Choquet}.

\subsection{Heaviside and Dirac Distributions}

We now further assume that a volume element $\omega$ is given on $M$.
Define the {\it Heaviside 0-form distributions} $\YPMzero$ by
\begin{equation}
\YPMzero[f] = \Yzero^\mpm[f] = \int_{\mpm} f \omega
\end{equation}
and the {\it Dirac 1-form distribution} $\Done$ by
\begin{equation}
\Done[V] = \Done^{\Sigma} [V]
         = \int_\Sigma \ip{V} \omega
\end{equation}
where we have given $\Sigma$ the orientation of $\partial M^-$.  Thus,
\begin{equation}
d\YPMzero = d\Yzero^\mpm = -\Done^{\partial\mpm} = \pm\Done^\Sigma = \pm\Done
\end{equation}
Since any regularly discontinuous tensor $T$ is locally integrable, the tensor
distribution $\hat{T}$ is defined by (\ref{Action}), which is equivalent to
\begin{equation}
\hat{T} = T^- \Yzero^- + T^+ \Yzero^+
\label{Master}
\end{equation}
where $T^\pm=T|_\mpm$.  Furthermore, since $\Disc{T}$ is locally integrable on
$\Sigma$, products such as $\Disc{T}\otimes\Done$ are defined.

\subsection{Differentiation of Regularly Discontinuous Tensors and the
		Associated Distributions}

Using (\ref{Master}) it is straightforward to compute the partial and exterior
derivatives of regularly discontinuous tensors.  For instance, if $f$ is a
regularly discontinuous function, then
\begin{equation}
\hat{f} = f^- \Yzero^- + f^+ \Yzero^+
\end{equation}
and if $f$ is piecewise $C^1$ we have, using (\ref{dID}),
\begin{equation}
d\hat{f} = df^- \Yzero^- + df^+ \Yzero^+ + \Disc{f} \Done
\label{dDisc}
\end{equation}
Extending the action of the exterior derivative to piecewise $C^1$
differential forms $\alpha$ by defining $d\alpha$ to be the piecewise $C^0$
form satisfying
\begin{equation}
(d\alpha)^\pm  = (d\alpha) |_\mpm = d (\alpha |_\mpm) = d(\alpha^\pm)
\label{Extension}
\end{equation}
allows this to be rewritten as
\begin{equation}
d\hat{f} = \widehat{\,df\,} + \Disc{f} \Done
\end{equation}

We turn now to covariant differentiation.  Assuming that there is a smooth
connection on $M$, we can again use (\ref{Master}) to determine the covariant
derivatives of piecewise smooth tensors.  In analogy with (\ref{Extension}),
we extend the covariant derivative to piecewise $C^1$ tensors $T$ by defining
the piecewise $C^0$ tensor $\Grad{X}T$ via
\begin{equation}
(\Grad{X}T)^\pm
  = (\Grad{X}T) |_\mpm
  = \Grad{X} |_\mpm (T |_\mpm)
  =: \Grad{X} T^\pm
\label{GradExt}
\end{equation}
It is now
straightforward to compute
\begin{eqnarray}
\Grad{X}(\YPMzero T^\pm)[U]
  &=& -(\YPMzero T^\pm) \Big[ \div(X) U + \GRAD{\pm}{X}U \Big] \nonumber\\
  &=& -\int_\mpm \langle T^\pm,\div(X)U+\GRAD{\pm}{X}U \rangle \, \omega
\end{eqnarray}
and
\begin{eqnarray}
(\YPMzero \Grad{X}T^\pm) [U]
  &=& \int_\mpm \langle \GRAD{\pm}{X}T^\pm,U \rangle \, \omega \nonumber\\
  &=& - \int_\mpm \langle T^\pm,\div(X)U+\GRAD{\pm}{X}U \rangle \, \omega
      \mp \int_\Sigma \langle T^\pm,U \rangle \, \ip{X}\omega
      \nonumber\\
  &=& \Grad{X}(\YPMzero T^\pm)[U] \mp \Done [\langle T^\pm,U \rangle X]
\end{eqnarray}
which verifies the product rule
\begin{equation}
\Grad{X}(\ypm T^\pm) = \ypm \Grad{X}T^\pm + \Grad{X}\ypm \otimes T^\pm
\end{equation}
For piecewise $C^1$ tensors $T$, $\Grad{X}T$ is regularly discontinuous, and
we can rewrite this as
\begin{eqnarray}
\widehat{\Grad{X}T}[U]
  &=& \Grad{X}\hat{T}[U]
      - \ip{X}\Done \big[\Disc{\langle T,U \rangle}\big]
\end{eqnarray}
which finally yields the expected formula for differentiating tensor
distributions associated with regularly discontinuous tensors, namely
\begin{equation}
\Grad{X}\hat{T} = \hat{\Grad{X}T} + \ip{X}\Done \otimes \Disc{T}
\label{Cute}
\end{equation}

\subsection{Regularly Discontinuous Connections}

A {\it regularly discontinuous connection} is a connection whose components
are regularly discontinuous (piecewise $C^0$).  The definition of the
covariant derivative on tensor distributions can be generalized to such
connections, provided its action is restricted to a subclass of the tensor
distributions.  We investigate here the consequences of taking this approach.

We would like to define the action on tensor distributions of the covariant
derivative corresponding to a regularly discontinuous connection.  There are
at least two ways to do this: 1) Use the component definition.  This succeeds
for all distributions such that the products of the connection symbols with
the component distributions are distributions.  While this does not work for
all distributions, the class for which it does work includes all distributions
associated with locally integrable tensors.  2) Use (\ref{Covariant}).  This
amounts to repeating the derivation in the previous subsection for this case,
and is outlined below.  In either approach, one recovers (\ref{Cute}).
Alternatively, (\ref{Cute}) itself could have been used to {\it define}
covariant differentiation of tensor distributions associated with piecewise
$C^1$ tensor fields.

Given (continuous) connections $\nabla^\pm$ on $\mpm$, we can construct the
(piecewise continuous) connection $\nabla$ on $M$ whose restrictions to $\mpm$
are just $\nabla^\pm$.  Given a piecewise $C^1$ tensor $T$ on $M$,
(\ref{GradExt}) again defines a piecewise $C^0$ tensor $\Grad{X}T$, which can
be expressed directly in terms of $\nabla^\pm$ as
\begin{equation}
(\Grad{X}T) |_\mpm = \GRAD{\pm}{X}T^\pm
\label{GradDef}
\end{equation}
Using this definition of $\Grad{X}U$, we can use (\ref{Covariant}) to define
the covariant derivative $\Grad{X}\Bold{T}$ of some tensor distributions
$\Bold{T}$. Specifically, we can do so for $\Bold{T}$ such that the map
$U\to\Bold{T}[\Grad{X}U]$ is a distribution.  Again, this includes all
distributions $\Bold{T}$ associated with locally integrable tensors.  If the
connection is regularly discontinuous, $\GRAD{\pm}{X}T$ and $\GRAD{\pm}{X}U$
are locally integrable, and the derivation of (\ref{Cute}) in the preceding
subsection can be used without modification.

\section{TENSOR DISTRIBUTIONS WITH DEGENERATE METRICS}
\label{WithoutMetric}

We consider both metrics $g$ which are {\it discontinuous} at a hypersurface
$\Sigma=\{\lambda=0\}$, and those which are {\it continuous} but have
vanishing determinant at $\Sigma$.  We assume in both cases that $g$ is
piecewise smooth, and that the pullback $h$ of $g$ to $\Sigma$ is the same
 from both sides and nondegenerate.  In particular, we assume that $\Sigma$ is
not null.  In both cases, $g$ takes the form
\begin{equation}
g = N \, d\lambda \otimes d\lambda + h_\lambda
\label{Metric}
\end{equation}
where $h_0=h$ and with $N$ being discontinuous (and nonzero) at $\Sigma$ in
one case and zero there in the other.

One way of treating the Levi-Civita connection determined by such a metric
would be to think of it as being singular on $\Sigma$, and perhaps to give it
the status of a distribution.  Instead, we wish to investigate the
consequences of thinking of it as being a regularly discontinuous function (in
the case of discontinuous metrics) or as the inverse of a smooth function
having a zero (in the continuous metric case).  This allows the covariant
derivative on tensor distributions previously defined to be used.

\subsection{Discontinuous Metrics}

For discontinuous metrics of the form (\ref{Metric}), one still has the
induced metric $h$ on $\Sigma$.  It thus seems natural to define the induced
volume element on $\Sigma$ to be $\sigma=\bar{*}1$ as before, which leads to
$\Dzero$ being well-defined.  Furthermore, the metric volume element on $M$ is
\begin{equation}
\omega = \sqrt{|N|} \, d\lambda \wedge \bar{*}1
\label{MetVol}
\end{equation}
which is discontinuous except when $N$ changes sign but not magnitude.  It is
remarkable that this special case of a {\it discontinuous signature-changing
metric} leads to a natural, continuous choice of volume element everywhere.

For discontinuous metrics which do not change signature,  there are two
different 1-sided volume elements $\omega^\pm$ and 1-sided unit normal 1-forms
$\mu^\pm$, satisfying
\begin{equation}
\omega^\pm = \mu^\pm \wedge \sigma
\end{equation}
One must now either choose a continuous, and hence non-metric, volume element
$\omega$ (e.g.\ by extending the metric volume element on one side), in which
case the volume elements $\sigma^\pm$ induced on $\Sigma$ by
$\omega=\mu^\pm\wedge\sigma^\pm$ will differ, or accept the existence of
separate metric volume elements $\omega^\pm$ on $\mpm$, which leads to two
different distributions $\Done^\pm=\pm d\YPMzero$.

Since the discontinuous metric $g$ admits nondegenerate 1-sided limits to
$\Sigma$, one can define the Levi-Civita connections $\nabla^\pm$ on the
manifolds-with-boundary $\mpm\cup\Sigma$.  In particular, $\nabla^\pm$ are
well-defined at $\Sigma$, although they will not in general agree there.
Thus, the connection defined by (\ref{GradDef}) is regularly discontinuous.
As discussed above, provided a smooth volume element has been chosen, partial
and exterior differentiation of tensor distributions can be defined.
Furthermore, covariant differentiation of distributions associated with
regularly discontinuous tensors can be defined by the usual component formula
or invariantly by (\ref{Covariant}).  Should the connection in fact be smooth,
of course, covariant differentiation of all tensor distributions can be
defined.  We emphasize that there is a natural choice of smooth volume element
for discontinuous, signature-changing metrics, namely (\ref{MetVol}).

\subsection{Continuous Metrics}

If however $N\AtSigma=0$, then the metric volume element (\ref{MetVol}) is
continuous, but zero on $\Sigma$.
\footnote{If we further assume that $dN\AtSigma\ne0$, then we have a typical
continuous signature-changing metric.  In this case, Kossowski and Kriele
\cite{KK} have shown that $N$ can always be chosen to be $\lambda$ near
$\Sigma$.}
Furthermore, there is no notion of unit normal vector to $\Sigma$.  One way
around this problem is again to work with a non-metric volume element, the
most obvious one being
\begin{equation}
\omega = d\lambda \wedge \sigma
\label{Volume}
\end{equation}
There is no difficulty defining tensor distributions with respect to this
volume element, using $\sigma$ as the natural volume element on $\Sigma$.

As we have shown, given any smooth volume element on $M$, we can define the
basic distributions, specifically the Heaviside and Dirac distributions, as
well as partial and exterior differentiation of distributions.  Consider now
covariant differentiation.  The Levi-Civita connection $\nabla^\pm$ on $\mpm$
for the metric (\ref{Metric}) is easily computed.  Note, however, that since
\begin{equation}
d\lambda (\nabla^\pm_{\partial_\lambda} \partial_\lambda)
  = {1\over2N} {\partial N\over\partial\lambda}
\label{BlowUp}
\end{equation}
which is singular on $\Sigma$,
$\nabla^\pm$ do not lead to a regularly discontinuous connection on $M$.  But
this means that it is not possible to define covariant differentiation of
distributions using (\ref{Covariant}).

As we have already pointed out, one can alternatively define covariant
differentiation of tensor distributions in terms of Christoffel symbols and
components.  For smooth connections, of course, both approaches agree.  For
continuous degenerate metrics, this approach requires giving meaning as a
distribution to the product of the Christoffel symbols with an arbitrary
distribution.  This in turn is essentially the same as defining the
distribution ${1\over\lambda}\Bold{D}$ for any distribution $\Bold{D}$.  This
can be done in many ways, as there are many distributions $\Bold{E}$
satisfying $\lambda\Bold{E}=\Bold{D}$; choosing one is equivalent to choosing
a preferred test function \cite{DeWitt}.  It is far from obvious to what
extent the resulting theory of tensor distributions, and in particular of
covariant differentiation thereof, depends on this choice.

\section{EXAMPLE WITH SIGNATURE CHANGE}
\label{Example}

A typical discontinuous signature-changing metric on ${\Bbb R}^2$ is
\begin{equation}
g_1 = \sgn(\tau) \, d\tau \otimes d\tau + a(\tau)^2 \, dx \otimes dx
\label{DiscEx}
\end{equation}
while a typical continuous signature-changing metric on ${\Bbb R}^2$ is
\begin{equation}
g_2 = t \, dt \otimes dt + a(t)^2 \, dx \otimes dx
\label{ContEx}
\end{equation}
where $a>0$.  These metrics can be identified away from $\Sigma=\{t=0=\tau\}$
via the transformation
\begin{equation}
\tau = \int_0^t \sqrt{|t|} \, dt
\end{equation}
but this fails to be a coordinate transformation at $t=0$, so that $t$ and
$\tau$ correspond to different differentiable structures on ${\Bbb R}^2$.  It
is instructive to note that in the ``$t$ differentiable structure'', in which
$t$ is an admissible coordinate, $d\tau=\sqrt{|t|}\,dt$ is zero at $\Sigma$,
whereas in the $\tau$ differentiable structure it is a basis 1-form and
nowhere zero.

\subsection{Heaviside and Dirac Distributions}

The metric volume elements determined by these metrics away from $\Sigma$ are
\begin{eqnarray}
\omega_1 &=& a \, d\tau \wedge dx \\
\omega_2 &=& a \, \sqrt{|t|} \, dt \wedge dx
\end{eqnarray}
which agree up to a coordinate transformation away from $\Sigma$.  However,
$\omega_2\AtSigma=0$, so that it is a degenerate volume element on ${\Bbb
R}^2$.  The pullback metric on $\Sigma$ is in both cases
\begin{equation}
h = a^2 \, dx \otimes dx
\end{equation}
with volume element
\begin{equation}
\sigma = a \, dx
\end{equation}

Even though the metric $g_1$ is discontinuous, not only the volume element
$\omega_1$ but also the unit normal 1-form $d\tau$ is continuous, and there is
no difficulty satisfying the relationship
\begin{equation}
\omega_1 = d\tau \wedge \sigma
\end{equation}
everywhere.  This leads to the distributions
\begin{eqnarray}
\YPMzero[f] &=& \int_{\pm\tau>0} f \, a \, d\tau \wedge dx \\
\Done[V]    &=& \int_{\tau=0} d\tau(V) \, a \, dx \\
\Dzero[f]   &=& \int_{\tau=0} f \, a \, dx
\end{eqnarray}
which satisfy
\begin{equation}
d\YPMzero = \Done = \Dzero \, d\tau
\end{equation}

For the continuous metric $g_2$, things are not so simple.  One possibility is
to choose the non-metric but nowhere degenerate volume element
\begin{equation}
\bar\omega_2 = a \, dt \wedge dx
\end{equation}
which leads to the distributions
\begin{eqnarray}
\bar\YPMzero[f] &=& \int_{\pm t>0} f \, a \, dt \wedge dx \\
\bar\Done[V]    &=& \int_{t=0} dt(V) \, a \, dx \\
\bar\Dzero[f] = \Dzero[f] &=& \int_{t=0} \, f \, a \, dx
\end{eqnarray}
These distributions satisfy
\begin{equation}
d\bar\YPMzero = \bar\Done = \bar\Dzero \, dt
\end{equation}

\subsection{Differentiation of Distributions}

Once an appropriate volume element is chosen, the definition of partial and
exterior differentiation of tensor distributions for signature-changing
metrics follows immediately from the general results in
Section~\ref{Differentiation}.  Turning to covariant differentiation, the
Christoffel symbols $\Gamma_i^\pm$ for the Levi-Civita connections
$\nabla_i^\pm$ associated with the metrics $g_i$ restricted to $M^\pm$ take
the form
\begin{eqnarray}
(\Gamma_2^\pm)^t{}_{tt} &=& {1 \over 2t} \qquad\quad~
  (\Gamma_1^\pm)^\tau{}_{\tau\tau} = 0 \\
(\Gamma_2^\pm)^x{}_{xt} &=& {a'\over a} \qquad\quad~
  (\Gamma_1^\pm)^x{}_{x\tau} = {\dot{a}\over a} \\
(\Gamma_2^\pm)^t{}_{xx} &=& -{aa' \over t} \qquad 
  (\Gamma_1^\pm)^\tau{}_{xx} = \mp a\dot{a}
\end{eqnarray}
in a coordinate basis, where derivatives with respect to $t$ ($\tau$) have
been denoted with a prime (dot).

As already noted, connections which are well-behaved on $\Sigma$ can be
extended via (\ref{Covariant}) to a connection on (a restricted set of) tensor
distributions.  The connections $\nabla_1^\pm$ associated with the
discontinuous signature-changing metric~(\ref{DiscEx}) each admit a
well-defined limit to $\tau=0$.  Thus, for discontinuous signature-changing
metrics, the Levi-Civita connection extends naturally to a definition of
covariant differentiation of distributions associated with regularly
discontinuous tensor fields.  Furthermore, this implies that the
distributional curvature (thought of as an operator on smooth tensors) can
always be defined.
\footnote{If $a$ satisfies the additional restriction $\dot{a}(0)=0$, the
connections on $\mpm$ can be extended to a smooth connection on all of $M$,
and thus to all tensor distributions, in a natural way, and the distributional
curvature corresponds to a (smooth) tensor.  This is not, however, a {\it
necessary} condition for the distributional part of the curvature tensor to
correspond to a locally integrable tensor.  The implications of this for
surface layers in signature-changing spacetimes has been discussed elsewhere
\cite{Einstein}.}

The connections $\nabla_2^\pm$ associated with the continuous
signature-changing metric (\ref{ContEx}) are singular at $t=0$, which means
that (\ref{Covariant}) can not be used to extend them to tensor distributions.
As discussed above, however, they can be so extended using the component
approach, provided one chooses a definition of ${1\over t}\Bold{D}$.  Having
done that, the Levi-Civita connection can be extended to all distributions.
To what extent this definition depends on the choice made, and whether the
resulting notion of distributional curvature does so, are open questions.

\subsection{Massless Klein-Gordon Equation}

Dray {\it et al.}\ \cite{PaperIII} postulate the massless Klein-Gordon
equation for (continuous) signature-changing metrics in the form
\begin{equation}
d\Bold{F} = 0
\label{Field}
\end{equation}
where

\begin{equation}
\Bold{F} = {*}d\Phi^- \Yzero^- + {*}d\Phi^+ \Yzero^+
\end{equation}
and where a regularity condition is assumed on ${*}d\Phi^\pm$, namely that
their 1-sided limits to $\Sigma$ exist.  In other words, ``${*}d\Phi$'' is
regularly discontinuous, where care must be taken to distinguish the different
Hodge dual operators on $\mpm$, both denoted by $*$.  Using (\ref{dID}), the
field equation (\ref{Field}) implies
\begin{equation}
\Yzero^\pm d{*}d\Phi^\pm = 0 = \Done \wedge \Disc{{*}d\Phi}
\end{equation}
which implies
\begin{eqnarray}
d{*}d\Phi^\pm &=& 0
\label{EqI} \\
\Sigma^* \Disc{{*}d\Phi} &=& 0
\label{EqII}
\end{eqnarray}
where $\Sigma^*$ denotes the pullback to $\Sigma$.  While the Heaviside
distributions $\Yzero^\pm$, and thus the distributional field $\Bold{F}$,
certainly depend on the choice of (nondegenerate) volume element used to
define tensor distributions, the resulting field equations (\ref{EqI}) and
(\ref{EqII}) do not.  Furthermore, since only limits of fields to $\Sigma$
occur in the crucial equation~(\ref{EqII}), and since the continuous and
discontinuous approaches are smoothly related away from $\Sigma$, one obtains
the same equations starting from a discontinuous metric.

\section{DISCUSSION}
\label{Discussion}

We began by constructing the Heaviside and Dirac tensor distributions
corresponding to an arbitrary volume element, and then showed in detail how to
differentiate tensor distributions.  We then applied our formalism to
manifolds with degenerate metrics, where the usual, metric-based definitions
fail.  In particular, we have shown that partial and exterior differentiation
can always be extended to tensor distributions; our extension depends only on
the volume element, and does not require a metric.

Our results concerning covariant differentiation are more subtle.  There are
situations in which the requirement that the connection be continuous is too
strong.  In particular, this would lead to the curvature tensor being at worst
discontinuous, thus eliminating the possibility of {\it distributional
curvature}.  We have therefore investigated the circumstances under which this
restriction can be weakened.

We can summarize our results in more informal language as follows.  Since the
component expansion of any discontinuous tensor on a smooth manifold consists
of discontinuous components multiplying smooth basis tensors, the usual rules
for a covariant derivative operator of course imply that
\begin{equation}
\Grad{X} (fT) = X(f)T + f \Grad{X}T 
\label{Usual}
\end{equation}
Thus, for any distribution associated with a regularly discontinuous tensor
field of the form
\begin{equation}
\Bold{T} = \bfy^- T^- + \bfy^+ T^+
\end{equation}
and using the action of vector fields implicit in (\ref{dDisc}) (i.e.\
obtained by contracting (\ref{dDisc}) with the vector field), we must have
\begin{equation}
\Grad{X}\Bold{T} = \bfy^- \Grad{X}T^- + \bfy^+ \Grad{X}T^+
                   + \Disc{T} \, \d(X)
\label{ConnTwo}
\end{equation}
With the convention that $\bfy^2=\bfy$ and $\bfy^+\bfy^-=0$, we can formally
rewrite this as
\begin{equation}
\nabla = \bfy^- \nabla^- + \bfy^+ \nabla^+
\label{Connection}
\end{equation}
and this formal prescription accurately reproduces our results provided we use
(\ref{Usual}) to differentiate the Heaviside functions.  Note that it is only
necessary here that $\nabla^\pm$ admit 1-sided limits to $\Sigma$.  If the
connection should be smooth, then $\nabla$ may be applied to any tensor
distribution, whereas if it is not then undefined products such as $\d\bfy$
would occur.  As a special case, our results can be applied to the case of a
continuous (nondegenerate) metric which is not differentiable, yielding the
usual formalism for the distributional curvature in that case.

One unusual feature of the connection (\ref{Connection}) is that even if
$\nabla^\pm$ are metric compatible, $\nabla$ will not be.  To see this, simply
compute
\begin{equation}
\nabla(\bfy^- g^- + \bfy^+ g^+) = \Disc{g} \otimes \d
\end{equation}
where the other two terms vanish due to the assumed metric-compatibility of
$\nabla^\pm$.  This result is unchanged even if $\nabla$ is assumed to be
smooth.  To avoid this, one might consider other extensions of $\nabla^\pm$,
which would differ from (\ref{Connection}) by the presence of a term
proportional to $\d$.  However, such a connection could only be applied to
smooth tensors and their associated tensor distributions; covariant
differentiation of discontinuous tensors using such a connection could not be
defined, as it would lead to unacceptable $\bfy\d$ terms.  Not only does this
mean that the distributional curvature could not be defined, but the metric
itself could not be differentiated, and there would be no way to establish
that such a connection is metric compatible in the first place!  If one
instead directly modifies (\ref{ConnTwo}), e.g.\ by removing the last term,
the argument leading up to (\ref{ConnTwo}) shows that such a connection would
not reduce to the correct action on discontinuous functions.  While we find
the introduction of a connection which is not metric compatible disconcerting,
we are forced to the conclusion that it is a necessary feature of covariant
differentiation for discontinuous metrics.  It is after all the unique
extension of $\nabla^\pm$ for which the question of metric-compatibility can
even be asked!

For continuous signature-changing metrics, the Levi-Civita connection can be
extended in several ways to all tensor distributions, depending on the
implicit choice of test function made when ``dividing a distribution by
zero''.  It is not clear what role this choice plays, and in particular it is
not clear what structure is independent of this choice.  For discontinuous
signature-changing metrics, it is remarkable that the Levi-Civita connection
can be naturally extended to tensor distributions, albeit only to a restricted
class and with the novel feature of metric incompatibility.  While this
distinction does not affect field equations such as the Klein-Gordon equation,
which can be expressed without using a connection, other field equations on
signature-changing manifolds, for instance Einstein's equations, may turn out
to be more naturally discussed using one of these approaches than the other.
This issue is further discussed in~\cite{Einstein}, where a variational
treatment of Einstein's equations using the discontinuous approach is given.

\newpage
\section*{ACKNOWLEDGMENTS}

It is a pleasure to thank George Ellis, David Hartley, Charles Hellaby,
Corinne Manogue, J\"org Schray, Robin Tucker, and Philip Tuckey for extensive
helpful discussions.  Further thanks are due the School of Physics \&
Chemistry at Lancaster University and the Department of Physics and
Mathematical Physics at the University of Adelaide for kind hospitality.  This
work was partially supported by NSF Grant PHY-9208494, as well as a Fulbright
Grant under the auspices of the Australian-American Educational Foundation.

\end{document}